\documentclass{pas}
\newcommand*{\lcdm}{\ensuremath{\Lambda}CDM}
\usepackage{multirow}
\usepackage{orcidlink}

\begin{document}

\lefttitle{Publications of the Astronomical Society of Australia}
\righttitle{J. Zhan and C. L. Reichardt}

\jnlPage{-}{-}
\jnlDoiYr{2025}
\doival{}

\articletitt{Research Article}

\title{Built-in precision: Improving cluster cosmology through the self-calibration of a galaxy cluster sample}

\author{\sn{Junhao} \sn{Zhan}\,\orcidlink{0009-0009-4000-4640} and \sn{Christian L.} \sn{Reichardt}\,\orcidlink{0000-0003-2226-9169}}

\affil{School of Physics, The University of Melbourne, Parkville, VIC 3010, Australia}

\corresp{Junhao Zhan, Email: junhzhan@student.unimelb.edu.au}

\citeauth{}


\begin{abstract}
We examine the potential improvements in constraints on the dark energy equation of state parameter $w$ and matter density $\Omega_M$ from using clustering information along with number counts for future samples of thermal Sunyaev-Zel'dovich selected galaxy clusters.
We quantify the relative improvement from including the clustering power spectrum information
for three cluster sample sizes from 33,000 to 140,000 clusters and for three assumed priors on the mass slope and redshift evolution of the mass-observable relation. 
As  expected, clustering information has the largest impact when
(i) there are more clusters and (ii) the mass-observable priors are weaker.
For current knowledge of the cluster mass-observable relationship, 
we find the addition of clustering information reduces the uncertainty on the dark energy equation of state, $\sigma(w)$, by factors of $1.023\pm 0.007$ to $1.0790\pm 0.011$, with larger improvements observed with more clusters. 
Clustering information is more important for the matter density, with $\sigma(\Omega_M)$ reduced by factors of $1.068 \pm 007$ to $1.145 \pm 0.012$. 
The improvement in $w$ constraints from adding clustering information largely vanishes after tightening priors on the mass-observable relationship by a factor of two. 
For weaker priors, we find clustering information improves the determination of the cluster mass slope and redshift evolution by factors of  $1.389 \pm 0.041$ and $1.340 \pm 0.039$ respectively. 
These findings highlight that, with the anticipated surge in cluster detections from next generation surveys, self-calibration through clustering information will provide an independent cross-check on the mass slope and redshift evolution of the mass-observable relationship as well as enhancing the precision achievable from cluster cosmology. 

\end{abstract}


\maketitle

\section{Intro}
\label{sec intro}

Observations of Type Ia supernovae \citep{Riess_1998,Perlmutter_1999} have provided the first evidence for the accelerating expansion of the Universe, which has been attributed to the existence of dark energy. 
Models of dark energy are generally classified by the dark energy equation of state parameter $w $ into three broad categories: a cosmological constant $\Lambda$ ($w=-1$), quintessence ($w>-1$) \citep{Quintessence} and phantom dark energy ($w<-1$) \citep{phantom}, each with distinct theoretical origins and consequences for cosmic evolution \citep{Dark_energy_review}. 
Dark energy plays a pivotal role in driving the late time expansion of the universe; therefore, accurately determining its equation of state is essential for understanding both the Universe's past and its future fate.

Surveys of galaxy clusters provide a powerful probe of dark energy. 
The growth of structures like galaxy clusters depends on gravity pulling matter together against the expansion of the Universe, thus 
an accelerating expansion slows down structure growth.
As peaks in the density field, samples of galaxy clusters can be used to test dark energy's effects on both the growth of structure and the expansion history of the universe.
 Thus the abundance of galaxy clusters as a function of mass and redshift offers a sensitive probe to cosmological parameters related to matter density and dark energy, such as $\Omega_M$, $S_8$, and $w$  \citep[e.g.,][]{Cluster_review}.

The thermal Sunyaev-Zel'dovich (tSZ) effect \citep{1972SZ}, which is due to the inverse Compton scattering of cosmic microwave background (CMB) photons off  hot electrons of the intracluster medium, is a promising technique to discover large numbers of galaxy clusters out to high redshifts. 
Because the energy gain of the scattered CMB photons in the non-relativistic regime is proportional to the initial photon energy, the tSZ signal is nearly redshift independent. 
Furthermore, the tSZ signal is proportional to the integrated electron pressure through the galaxy cluster, which is expected to be a low-scatter proxy for the cluster mass \citep[e.g.,][]{Kay_2012}. 
The combination offers the prospect of nearly mass-selected cluster samples out to their redshift of formation, which has great potential for cluster cosmology \citep{carlstrom02}.

The precision of cosmological tests from galaxy cluster abundances depends on how well the cluster masses can be inferred from the observed signals, and thus requires a precisely calibrated mass-observable relation. 
 One common approach is to calibrate the mass-observable relation via external weak lensing data \citep[e.g.,][]{Reichardt_2013,Dietrich2019,bocquet2024}, but acquiring such data can be expensive and observationally demanding especially for high redshift clusters. 
 An alternative approach is the use the spatial clustering of the sample itself to help calibrate the inferred cluster masses, with this `self-calibration' method first proposed by \citet{MM04}, and developed further by others \citep[e.g.,][]{Lima_2004,Lima05,Lima_2007,Hu_Cohn_2006,Wu_2008,Oguri09}. 
 The self-calibration method involves dividing the cluster catalogue into redshift bins and, within each bin, using both the cluster abundance and the redshift averaged clustering power spectrum to constrain cosmology and the mass-observable relation simultaneously. 
 
Extracting useful spatial clustering information requires reaching a sufficiently high density of galaxy clusters in the sample. 
The predictions in the early 2000s were predicated on large cluster samples, which would be reached at higher survey noise levels when $\sigma_8\sim 1$ was plausible. 
However, current observations favor $\sigma_8\sim 0.8$, which leads to an order of magnitude reduction in expected cluster sample sizes.  
As a result, only now are CMB surveys such as South Pole Telescope (SPT) \citep{spt_3G}, Simons Observatory (SO) \citep{SO_2019,SO_2025},  and potential future stage 4 surveys like the proposed CMB-S4 \citep{cmbs4sciencebook} or CMB-HD \citep{CMBHD} expecting to provide catalogues with $10^4$ -- $10^5$ tSZ-selected galaxy clusters.  
The large number of clusters expected from these surveys make this the ideal time to revisit the potential of self-calibration as an effective tool for enhancing galaxy cluster cosmology and constraining dark energy.

In this paper we forecast expected improvements to constraints on the dark energy equation of state parameter $w$ and matter density $\Omega_M$ that come from adding clustering information to cluster abundance measurements. 
We also explore how the results depend on the number of clusters, redshift uncertainties, and how well characterised the tSZ-mass scaling relation is from other data as weak lensing measurements and X-ray mass estimates.

This paper is organised as follows. 
We describe the simulated cluster catalogues used in Section 2, before presenting the estimators for the angular power spectrum and covariance in Section 3. 
The modeling of cluster abundance, clustering and the likelihood function is described in Section 4.
In Section 5, we present our results before concluding in Section 6.

\section{Simulated Cluster catalogues}
\label{sec 2}

The forecasts in this work are derived from simulated galaxy cluster catalogues extracted from the  HalfDome simulation suite. 
In this section, we introduce the HalfDome simulations, from which lists of clusters are obtained, along with their true redshifts, masses and positions. 
We then describe  the treatment of redshift uncertainties and the tSZ signal, and finish by summarising the cluster sample selection criteria for each of the three cluster samples used in this analysis.

\subsection{HalfDome Simulation}
\label{sec HalfdomeSim}

We draw cluster samples from the HalfDome cosmological N-body simulations  \citep{halfdome}. A key advantage of the HalfDome simulations is the availability of 11 independent full-sky realisations of the fiducial cosmology, each with a full-sky catalogue of clusters up to $z = 4$. Multiple catalogues allow more robust statistical tests on the effect of using self-calibration for the cluster sample. 
Additionally, the HalfDome cluster masses have been calibrated to match the  $M_{\rm 200m}$ mass definition of the Tinker mass function \citep{Tinker_2008}. This minimises potential parameter biases due to a mismatch halo mass function between the catalogue and theory predictions in the Markov Chain Monte Carlo (MCMC). We refer readers to \cite{halfdome} for more information on the simulation and catalogues.

The HalfDome simulations are based on a \lcdm{} cosmology with the Planck 2015 cosmological parameters \citep{Planck2015}: $h = 0.6774$, $\Omega_M = 0.3089$, $\Omega_b = 0.0486$, $\sigma_8 = 0.8159$, $n_s = 0.9667$, $M_\nu = 0$, and $w = -1$.

\subsection{Redshift uncertainties}
\label{sec: redshifts}

We introduce errors on the true cluster redshifts from the HalfDome simulations to mimic redshift measurement uncertainties.  
For each cluster, we draw an observed redshift from $\mathcal{N}(z,\sigma_z^2)$ where $z$ is the true redshift from HalfDome and the error $\sigma_z$ is defined as:
\begin{equation}
    \label{z error}
    \sigma_z = \sigma_{z,0}(1+z) .
\end{equation}
We consider three cases: $\sigma_{z,0}=0$ (no error), as well as $\sigma_{z,0}=0.01$ and 0.02.
No difference is seen in the resulting cosmological constraints between these cases. 
We use $ \sigma_{z,0} = 0.02$ for all numbers presented in this work.
We neglect the possible effect of peculiar velocities, as the impact on these scales has been shown to be minor \citep{Romanello_2025}.

\subsection{tSZ signal}
\label{sec: obs errors}

To simplify the forecasting, we use a tSZ-mass scaling relation to assign tSZ signals to the HalfDome catalogues of collapsed matter haloes rather than running a multi-frequency cluster finding algorithm on the HalfDome maps. 
We assign mock tSZ signal-to-noise estimates to each cluster, using a mass-S/N relationship used by recent ground-based tSZ cluster surveys, from the South Pole Telescope and Atacama Cosmology Telescope. 
Following \cite{bocquet2024} and \cite{SPT_Deep_catalogue_2025}, we model the relationship of each cluster's unbiased significance $\zeta$ to the cluster mass and redshift as 
\begin{equation}
    \zeta = A_{sz} \biggr(\frac{M_{\rm 200m}}{3 \times 10^{14} h^{-1}M_{\odot}}\biggr)^{B_{sz}} \biggr(\frac{H(z)}{H(0.6)}\biggr)^{C_{sz}} e^{\mathcal{N} (0,\sigma_{\ln \zeta}^2)} ,
\end{equation}
where $A_{sz}$ is an overall mass calibration, $B_{sz}$ encodes the mass dependence, $C_{sz}$ the redshift dependence, and the final exponential term introduces a log-normal scatter with variance $\sigma_{\ln \zeta}^2$.  
We generate tSZ significances,  $\zeta$, for each HalfDome halo using the central values from \cite{bocquet2024}: 
$\ln{A_{sz}}=0.69; B_{sz}=1.73; C_{sz}=0.74; \sigma_{\ln \zeta} = 0.21$.
Note that we use mass $M_{\rm 200m}$, which is provided by the HalfDome catalogues, instead of $M_{\rm 200c}$ without adjusting the central values from  \cite{bocquet2024}. 
Mass conversions can potentially introduce systematic biases \citep{Euclid_halo}.

\subsection{Sample selection}
\label{subsec:sampleselect}

We assume the cluster catalogues are selected from a large-area Chilean survey, like SO or CMB-S4, covering the declination range $-68^\circ$ to $+27^\circ$. 
After applying a galactic cut and a $2^\circ$ border apodization, the sky fraction is $f_{sky}=0.60$, consistent with expectations from SO \citep{SO_2025} and CMB-S4 \citep{cmbs4sciencebook}. 
We extract clusters located within this survey area for each of the 11 HalfDome realisations, and restrict each catalogue to the subset of clusters with an observed redshift between 0.2 and 2.

We aim for three final cluster sample sizes of 33,000 clusters, 70,000 clusters or 140,000 clusters to match forecasts for experiments like SO and CMB-S4. 
Forecasts for the SO enhanced LAT survey predict a sample of $\sim$33,000 clusters  \citep{SO_2025}. 
The proposed CMB-S4 survey has been predicted to find up to $\sim 140,000$ clusters \citep{cmbs4sciencebook}, while the 70,000 cluster sample splits the difference. 
The estimated detection significance in Section~\ref{sec: obs errors}, which is based on a past survey, is naturally an underestimate of the expected detection significance that will be achieved by these lower-noise, future surveys. 
Thus, we set a threshold on signal-to-noise $\zeta$ based on the desired sample size, which corresponds to thresholds of $\zeta > 1.43$,  $\zeta > 0.90$, and  $\zeta > 0.57$ for the 33,000 clusters, 70,000 clusters and 140,000 cluster samples respectively.

\section{Clustering Power Spectrum estimation}
\label{sec Observed Angular Power Spectrum}

We calculate the angular power spectrum for clusters in each redshift/significance bin using a catalogue-based pseudo-$C_\ell$ method \citep{Wolz_2025} as implemented by the publicly available NaMaster code \citep{NaMaster}. 
The angular power spectrum is the Hankel transform of the two-point correlation function, and thus encodes information on the excess probability to find galaxy clusters separated by some angular separation compared to a random distribution. 
We briefly summarise the technique here, and  refer the reader to \cite{NaMaster} and \cite{Wolz_2025} for further details.

\subsection{Angular Power Spectrum Estimator}

\begin{figure}[htbp]
\centering
\includegraphics[width=\columnwidth]{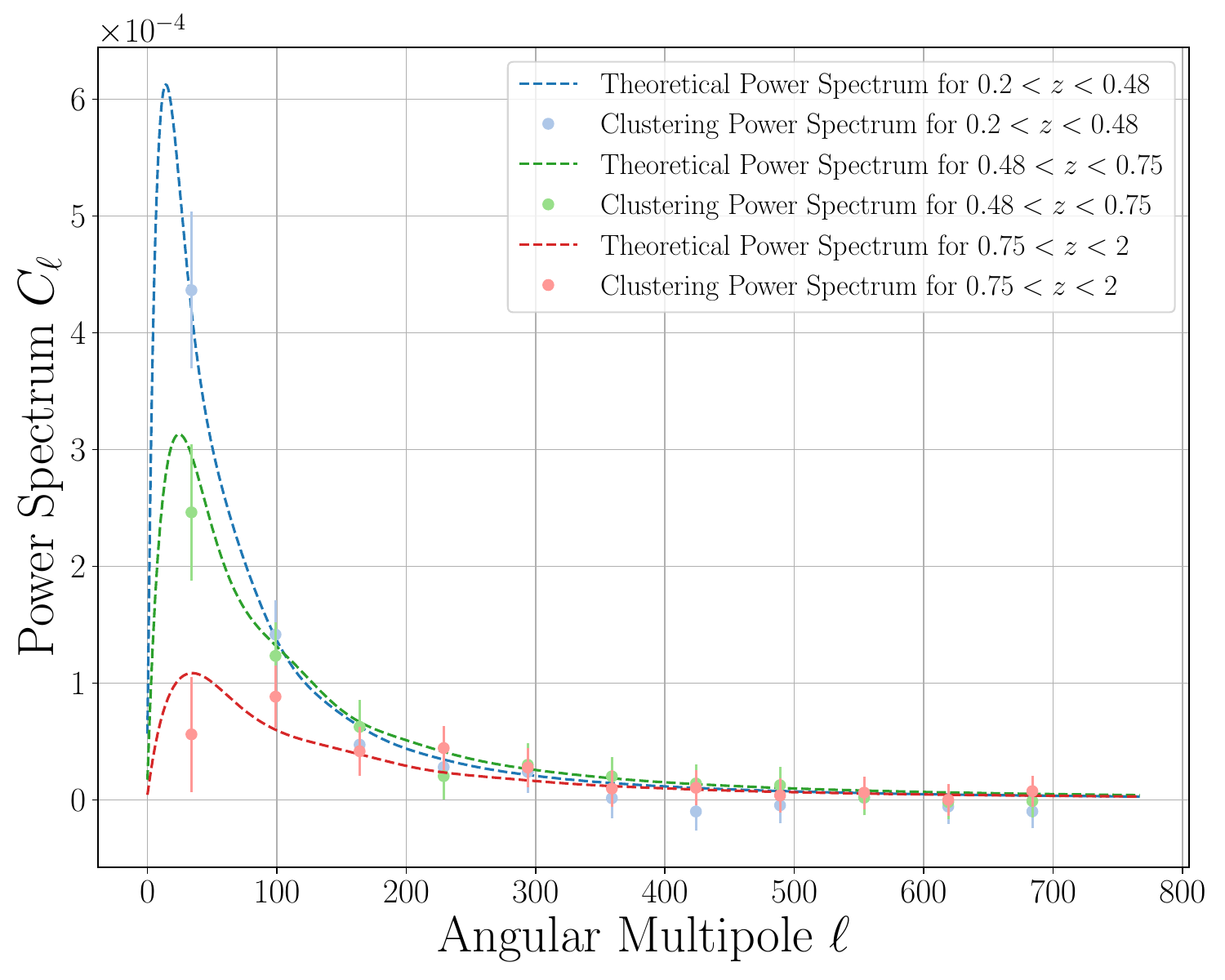}
\caption{\label{fig:PS} Measured bandpowers compared to theoretical predictions across three redshift bins for one realisation of the 33k cluster sample. 
The impact of clustering is most significant at low angular multipoles ($\ell < 300$), and drops towards zero at higher multipoles. 
The amplitude in the power spectrum increases at lower redshifts due to the continued growth of structure over time. }
\end{figure}

We use the NaMaster software package to calculate the clustering power spectrum for the clusters in each redshift or observable-redshift bin. 
We first extract the subset of $N_{bin}$ clusters that fall within a given bin. 
A second, larger random catalogue with $50\times N_{bin}$ objects at random positions is also generated. 
Both catalogues are weighted by the apodized mask described in Section~\ref{subsec:sampleselect}, pixelated at \texttt{nside} $= 256$. 
The mask and both catalogues are passed to the NaMaster \texttt{NmtFieldCatalogClustering} functions to produce a binned estimate of the clustering power spectrum that has been corrected for shot-noise and the selection function of the mask. 
The measured power spectra are binned into 11 angular multipole bins with equal widths of  $\Delta\ell = 65$, which we refer to as bandpowers. 
Figure~\ref{fig:PS} shows the measured bandpowers in three redshift bins for one 33k cluster sample.

We have tested binning the cluster sample only in redshift or in both redshift and SZ signal-to-noise space. 
We find no significant difference in the resulting parameter constraints between the two approaches. 
As there is some computational advantage to having fewer bins, the remainder of this work only reports results from binning in redshift only.
A dynamical redshift bin size is used for each realisation based on two requirements: each bin must contain at least $10,000$ clusters and a minimum redshift width of $\Delta z_{photo} \ge 0.2$.
The latter is set to ensure the covariance between redshift bins is small, 
allowing the covariance matrix to be treated as diagonal. 
These criteria result in three redshift bins for the 33,000 cluster samples, five redshift bins for the 70,000 cluster samples, and six redshift bins for the 140,000 cluster samples. 

\subsection{Covariance Matrix}
\label{sec Bootstrapping}
\begin{figure}[htbp]
\centering
\includegraphics[width=\columnwidth]{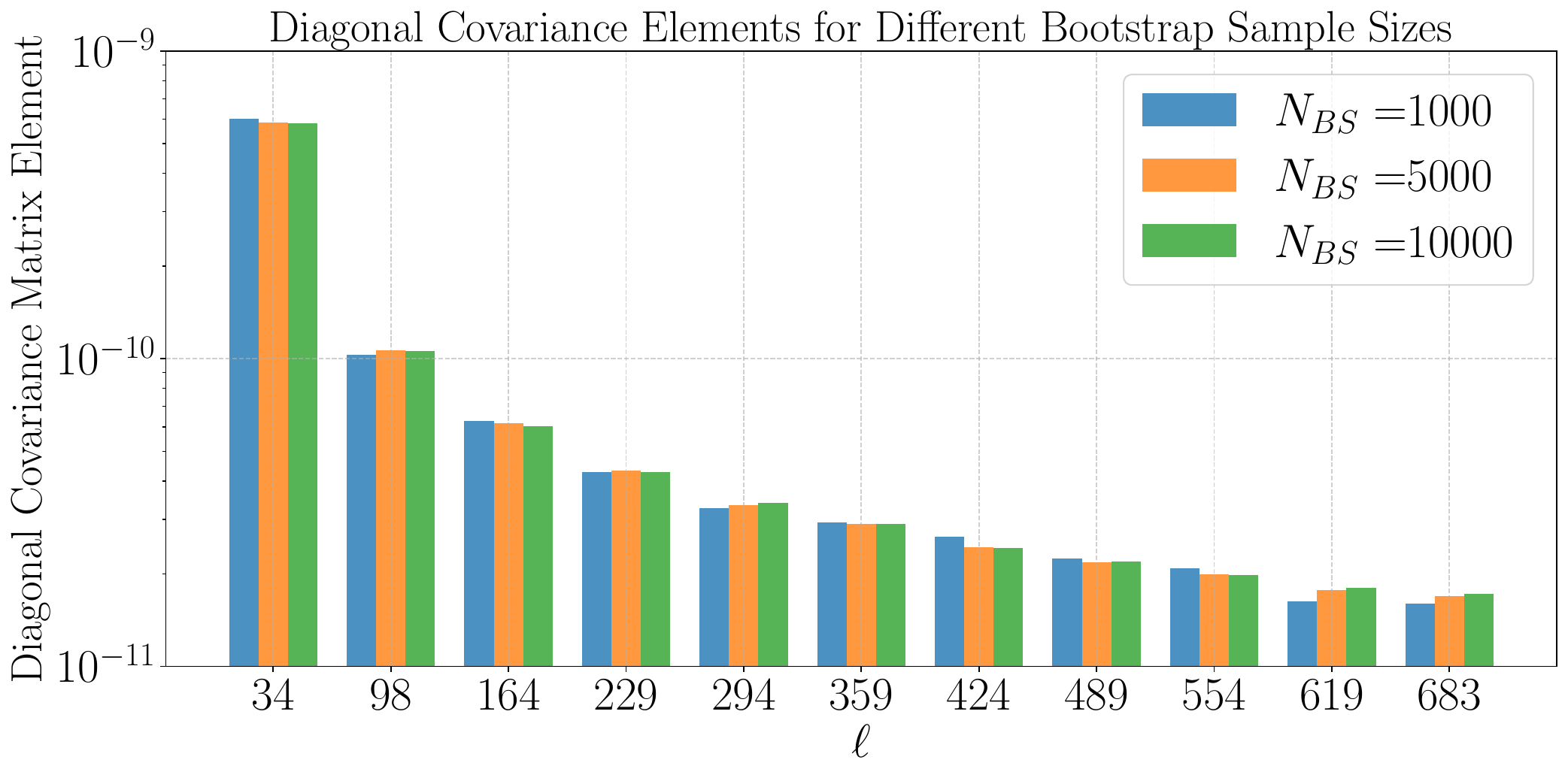}
\caption{The convergence of the covariance matrix estimate as $N_{BS}$ is increased. The covariance is well-estimated with $N_{BS}= 5000$ samples, with $<5\%$ shifts when doubling the number of draws to $N_{BS}= 10000$.  \label{fig:N_bs test}}
\end{figure}

We use bootstrap resampling to estimate the covariance matrix of the bandpowers.
We randomly draw $N_{BS}$ subsets of the clusters in each redshift bin $i$ and compute the bandpowers $C_b$ of each subset $j$. 
The covariance matrix $\hat{\Sigma}_{bb',i}$ can be estimated as:
\begin{equation}
    \hat{\Sigma}_{bb',i} = \frac{1}{N_{BS}-1} \sum_{j=1}^{N_{BS}}(C^j_b- \bar{C_b})(C^j_{b'}- \bar{C_{b'}}).
\end{equation}
We have tested the convergence of the covariance matrix estimation with the number of bootstraps $N_{BS}$, and show the results in Figure \ref{fig:N_bs test}. 
The results converge by $N_{BS} = 5000$ random draws, with an increase to $N_{BS} = 10,000$ yielding less than 5\% shifts in covariance elements. 
For the main results of this paper, we use $N_{BS} = 5000$ random draws.

To handle the bias incurred by uncertainty in the covariance matrix estimate, we apply the Hartlap correction \citep{Hartlap_2006}:
\begin{equation}
    [\Sigma]_{bb',i}^{-1} = \frac{N_{BS} - N_p- 2}{N_{BS}-1} [\hat{\Sigma}]_{bb',i}^{-1} .
\end{equation}
The correction is very small at $ \approx$ 0.2\% for the number of bootstrap resamples $N_{BS}$ and bandpowers $N_p$ used in this work.

\section{Parameter estimation}
\label{sec 4}

We forecast cosmological parameter constraints  using MCMC sampling with the cobaya package \citep{Cobaya}. 
In all cases, we run 11 chains, one per HalfDome realisation, and stop sampling at a Gelman-Rubin statistic \citep{Gelman-Rubin} of $\hat{R}-1 < 0.01$. The parameter space includes wCDM cosmological parameters and four scaling relation parameters. 
As the cluster data do not constrain all seven wCDM parameters, we fix the optical depth $\tau$ which does not affect the cluster counts, and set tight Planck 2018 inspired priors \citep{Planck2018}  on the Hubble parameter $h$, scalar spectral index $n_s$, and baryon density $\Omega_b h^2$.
We set non-informative uniform priors on the final three cosmological parameters, the matter density $\Omega_M$, clustering amplitude $S_8 = \sigma_8 \sqrt{\Omega_M/0.3}$, and dark energy equation of state parameter $w$. 
The cosmological parameters and priors are listed in the top panel of Table~\ref{tab:Priors}.

Uncertainty in the mass-observable scaling relation is the major systematic uncertainty in cluster cosmology, motivating self-calibration techniques. 
We adopt scaling relation priors based on the quoted uncertainties from a recent SPT cluster cosmology work \citep{bocquet2024} for $\ln{A_{sz}}$, $B_{sz}$ and $C_{sz}$. 
In testing we find the addition of clustering information primarily affects the mass slope and redshift evolution parameters, $B_{sz}$ and $C_{sz}$. 
To explore the role of external mass information, we vary the prior widths on these two parameters by scaling their quoted uncertainties by a constant factor $\Delta$.
We report results for $\Delta \in [0.5, 1, 2]$, which we refer to, respectively, as the optimistic, baseline and pessimistic cases. 
Following \cite{bocquet2024}, we also set a lower bound on the scatter of $\sigma_{\ln \zeta}>0.05$ to avoid numerical instability. 
We have tested imposing a Gaussian (instead of uniform) prior on $\sigma_{\ln \zeta}$ based on the posterior in \cite{bocquet2024}, and have  found this has no significant effect on the relative improvement due to the addition of clustering information.

\begin{table}[htbp]
\centering
\caption{Summary of the variables and priors used in the analysis. $\mathcal{U}(\text{min, max})$ stands for a uniform prior, while $\mathcal{N}(\mu,\sigma^2)$ stands for a Gaussian prior with mean $\mu$ and standard deviation $\sigma$. 
The central values are set to the values used by the HalfDome simulations (see Section~\ref{sec HalfdomeSim}).
\label{tab:Priors} }
\resizebox{\columnwidth}{!}{
\begin{tabular}{llr}
\hline
\textbf{Parameters} & Description & Prior \\
\hline
\multicolumn{3}{l}{\textbf{Cosmology}} \\
$\Omega_M$     & Matter density & $\mathcal{U}(0.1,0.9)$ \\
$S_8$     & Clustering amplitude & $\mathcal{U}(0.3,1.5)$ \\
$w$     & Dark energy equation of state parameter & $\mathcal{U}(-3,-0.33)$ \\
$h$     & Hubble parameter & $\mathcal{N}(0.6774,0.0081^2)$ \\
$n_s$     & Scalar spectral index & $\mathcal{N}(0.9667,0.0038^2)$ \\
$\Omega_b h^2$     & Baryon density & $\mathcal{N}(0.0223,0.00014^2)$ \\
\hline
\multicolumn{3}{l}{\textbf{Mass observable parameters}} \\
$\ln{A_{sz}}$     & Amplitude & $\mathcal{N}(0.69,0.06^2)$ \\
$B_{sz}$     & Mass slope & $\mathcal{N}(1.73,(0.04\Delta)^2)$ \\
$C_{sz}$     & Redshift evolution & $\mathcal{N}(0.74,(0.11\Delta)^2)$ \\
$\sigma_{\ln \zeta}$     & Intrinsic scatter & $\mathcal{U}(0.05,1)$ \\
\hline
\end{tabular}
}
\end{table}

\subsection{Galaxy cluster abundances}

We evaluate the likelihood of the observed cluster abundances for a given model  using the Cash C-statistic \citep{Cash_C}:
\begin{equation}
    \label{binned Cash C}
    \ln{\mathcal{L}} = o_i \ln(e_i) - E .
\end{equation}
Here $E$ is the  total number of clusters predicted for the catalogue while $o_i$ and $e_i$ are respectively
the observed and expected number of clusters in each tSZ signal-to-noise and redshift bin $i$. 
The bin width is $\Delta z_{o} = 0.01$ in redshift and $\Delta \ln \zeta = 0.05$ in logarithmic tSZ signal-to-noise.
The expected number in each bin is calculated by binning the number density in observable space:
\begin{align}
\label{eq: dN/dlnzetadzp}
\frac{d^2N(\ln \zeta, z_o | \mathbf{p})}{d\ln{\zeta}\,dz_o} = 
& \int_{0}^{\infty} \int_{0}^{\infty} d\ln M \, dz \,
P(\ln \zeta, z_{o}|\ln M, z,\mathbf{p}) \nonumber \\
& \times \frac{d^2N(M,z|\mathbf{p})}{d \ln M\, dz} \, ,
\end{align}
calculated from the integral over the true cluster mass $M$ and redshift $z$. 
Here the function $P(\ln \zeta, z_{o}|\ln M, z,\mathbf{p})$ is the probability to observe $\ln \zeta$ and $z_{o}$, 
$\mathbf{p}$ is a vector of all the cosmological and mass-observable parameters, 
and $\frac{d^2N}{d \ln M dz}$ is the differential number of clusters in true cluster mass and redshift space. 
We use the Tinker mass function \citep{Tinker_2008}, calibrated for the $\rm 200m$ mass definition.

\subsection{Clustering power spectrum}

The likelihood of observing clustering bandpowers $C_{b,j} $ in a redshift bin $j$ can be written as 
\begin{equation}
    \ln{\mathcal{L}} = - \frac{1}{2} \times 
    (C_{b,j} - C^{th}_{b,j}) [\Sigma^{-1}]_{bb',j} (C_{b',j} - C^{th}_{b',j}) .
\end{equation}
Here, $\Sigma^{-1}_j$ is the inverse of the covariance matrix for redshift bin $j$ estimated in Section~\ref{sec Bootstrapping}, $C_{b,j}$ are the observed bandpowers for that redshift bin, and $C^{th}_{b,j}$ is the predicted binned spectrum which is estimated by applying the bandpower window function to the theoretical spectrum. 

The theoretical power spectrum is estimated using the Limber approximation \citep{Limber_Approx}.
Under this assumption, the clustering power spectrum for the observed redshift bin $j$ can be expressed \citep{Romanello_2025} as: 
\begin{align}
\label{Limber Approximation eq}
C^{\text{th}}_{\ell}(\mathbf{p}) = \int^{\infty}_{0} dz \, 
&\, b^2_{\text{eff}}(> \zeta_{\text{cut}}, z| \Delta z_o, \mathbf{p}) \,
n_j^2(z| \Delta z_o, \mathbf{p}) \nonumber \\
&\times P_m\left( \frac{\ell + 0.5}{\chi}, z | \mathbf{p} \right)
\left( \frac{d^2V(z|\mathbf{p})}{d\Omega\, dz} \right)^{-1}
\end{align}
where $\chi$ is the comoving distance, $\frac{d^2V}{d\Omega dz}$ is the comoving volume element and $P_{m}$ is the matter power spectrum. 
The density kernel $n_j$ gives the probability distribution of true redshift for sources in the bin given the cosmology and observed redshift range, $\Delta z_o$.
The effective linear bias  $b_{eff}$  marginalised to signal space is calculated from the Tinker model  \citep{Tinker_2010} using the approach in \cite{MM04}.

The mass function, halo bias and matter power spectrum are computed using the publicly available code, Colossus \citep{Colossus}. 
For computational speed, this code uses the Hu \& Eisenstein approximation \citep{Eisenstein_Hu_approx} which tends to overestimate matter power spectrum at small scales. 
While this overestimate should bias  both the abundance and clustering constraints, we do not expect the bias to affect the relative improvement from adding clustering information.

\section{Results}
\label{Results}
We find that the addition of clustering information to galaxy cluster number counts leads to  modest reductions of  tens of percents in the allowed parameter volume. 
Clustering information leads to the largest reduction in parameter volume when we have (i) more clusters and (ii) less accurate mass estimates. 
For the baseline mass-observable priors, which are based on recent cluster cosmology results, the  inclusion of clustering measurements reduces
the allowed parameter volume by factors of 1.2, 1.3 and 1.4 for the 33k, 70k and 140k cluster samples respectively.
With weaker external priors (the pessimistic case). the reductions become more significant at 1.5, 1.7 and 2.0 respectively. 
Conversely, for stronger external priors (the optimistic case), clustering information has a smaller impact, with reductions by factors of 1.1 - 1.3 for the three sample sizes. 
While these improvements  are certainly modest across a multi-dimensional parameter space, we note that the inclusion of clustering information provides an avenue for an independent (though statistically weaker) cross-check of the cluster mass estimates.
We also stress that the improvements due to measuring the clustering of galaxy clusters are essentially free --- calculating the clustering power spectra does not require any additional observations beyond those needed for cluster number counts.

In the next sections, we look at how these parameter volumes reductions project onto the 1D constraints on individual parameters. 
We first look at the results for the set of wCDM cosmological parameters, followed by the mass-observable scaling relation parameters.

\subsection{Cosmological Parameters}
\label{sec: Cosmological Parameter}

We find that the addition of clustering information makes the largest difference in measuring the matter density  $\Omega_M$, with the dark energy equation of state parameter $w$ showing the second-largest improvement. 
We see little change to the 1D constraints on other parameters. 

The uncertainty on the dark energy equation of state parameter $w$ is reduced by a factor of $1.079\pm 0.011$ for the 140k cluster sample under the baseline mass-observable priors. 
This reduction drops to a factor of $1.042\pm 0.009$ for the smaller 70k cluster sample, and to a factor of $1.023\pm 0.007$ for the smallest cluster sample. 
With more information on the mass-observable relationship (the optimistic prior case), the improvements nearly vanish for the dark energy equation of state parameter. 
We find the uncertainty on $w$ is reduced by only a factor of $1.021\pm 0.006$ for the largest sample when clustering information is combined with the optimistic priors. 
Conversely with the weaker priors of the pessimistic case, the addition of clustering information yields sizeable improvements by a factor of up to $1.196\pm 0.025$ to $\sigma(w)$ for the largest cluster sample. 
The improvement factors for $\sigma(w)$ in each case are tabulated in the top section of Table~\ref{tab:improvements}.
\begin{table*}[htbp]
\centering
\caption{The relative improvement factors for the parameters of interest, defined as $\frac{\sigma_{\rm abundance}}{\sigma_{\rm abundance+clustering}}$, for the three cluster sample sizes and three prior options. 
Clustering information contributes to larger improvements for larger cluster samples with weaker scaling relation priors.  \label{tab:improvements}}
\resizebox{0.8\textwidth}{!}{
\begin{tabular}{|l|c|c|c|}
\hline
\textbf{Sample} & \multicolumn{3}{c|}{\textbf{Multiplication factor ${\Delta}$ of Priors for Mass-Observable Scaling Parameters}} \\
\cline{2-4}
& \quad Optimistic ($\Delta = 1/2$) \quad  & \quad \quad \quad  Baseline ($\Delta = 1$) \quad \quad \quad & Pessimistic ($\Delta = 2$) \\
\hline
\multicolumn{4}{|c|}{Relative Improvement in \textbf{${\sigma(w)}$}} \\
\hline
33k     & $1.014 \pm 0.005 $ & $1.023 \pm 0.007 $ & $1.091 \pm 0.014 $  \\
70k     & $1.005 \pm 0.006 $ & $1.042 \pm 0.009 $ & $1.118 \pm 0.014 $ \\
140k       & $1.021 \pm 0.006 $ & $1.079 \pm 0.011 $ & $1.196 \pm 0.025 $  \\
\hline
\multicolumn{4}{|c|}{Relative Improvement in \textbf{${\sigma(\Omega_M)}$}} \\
\hline
33k    & $1.058 \pm 0.005 $ & $1.068 \pm 0.007 $ & $1.112 \pm 0.009 $  \\
70k    & $1.058 \pm 0.008 $ & $1.082 \pm 0.007 $ & $1.160 \pm 0.019 $  \\
140k    & $1.085 \pm 0.008 $ & $1.145 \pm 0.012 $ & $1.288 \pm 0.040 $  \\
\hline
\multicolumn{4}{|c|}{Relative Improvement in \textbf{${\sigma(S_8)}$}} \\
\hline
33k    & $1.018 \pm 0.005 $ & $1.034 \pm 0.004 $ & $1.074 \pm 0.006 $  \\
70k       & $1.030 \pm 0.004 $ & $1.046 \pm 0.004 $ & $1.091 \pm 0.006 $  \\
140k      & $1.054 \pm 0.007 $ & $1.078 \pm 0.007 $ & $1.120 \pm 0.015 $ \\
\hline
\multicolumn{4}{|c|}{Relative Improvement in \textbf{${\sigma(B_{sz})}$}} \\
\hline
33k    & $1.020 \pm 0.010 $ & $1.046 \pm 0.069 $ & $1.114 \pm 0.015 $  \\
70k    & $1.043 \pm 0.010 $ & $1.100 \pm 0.012 $ & $1.239 \pm 0.027 $  \\
140k    & $1.079 \pm 0.010 $ & $1.192 \pm 0.019 $ & $1.389 \pm 0.041 $  \\
\hline
\multicolumn{4}{|c|}{Relative Improvement in \textbf{${\sigma(C_{sz})}$}} \\
\hline
33k    & $1.005 \pm 0.006 $ & $1.031 \pm 0.007 $ & $1.129 \pm 0.021 $ \\
70k    & $1.025 \pm 0.004 $ & $1.089 \pm 0.006 $ & $1.224 \pm 0.028 $  \\
140k    & $1.041 \pm 0.005 $ & $1.142 \pm 0.014 $ & $1.340 \pm 0.039 $ \\
\hline
\end{tabular}
}
\end{table*}

The addition of clustering information yields the most significant improvements to  constraints on the matter density $\Omega_M$. 
Even with strong external mass information (the optimistic prior case), clustering information reduces $\sigma(\Omega_M)$ by a factor of $1.085\pm 0.008$ for the 140k cluster sample, of order four times the reduction seen for $\sigma(w)$ in this case. 
The improvements are more significant for the other prior cases, as expected. 
We find the uncertainty on the matter density is reduced by a factor of $1.145\pm0.012$ for the baseline prior case with a 140k cluster sample, jumping to a factor of $1.288\pm 0.040$ with the pessimistic priors. 
The second section of Table~\ref{tab:improvements} has the results for $\Omega_M$.

We find clustering information reduces in the uncertainty on  $S_8  = \sigma_8 (\Omega_M/0.3)^{0.5}$ but not on $\sigma_8$ itself. Since the uncertainty in $\sigma_8$ does not improve,
the improvement for $S_8$  is straightforwardly explained from the improved measurement of $\Omega_M$ enabled by clustering information. 
The $S_8$ numbers are listed in the third section of Table~\ref{tab:improvements}, and are of order half the size of the reductions for $\sigma(\Omega_M)$.

\subsection{Understanding the constraints}

\begin{figure*}[htbp]
\centering
\includegraphics[width=1\textwidth]{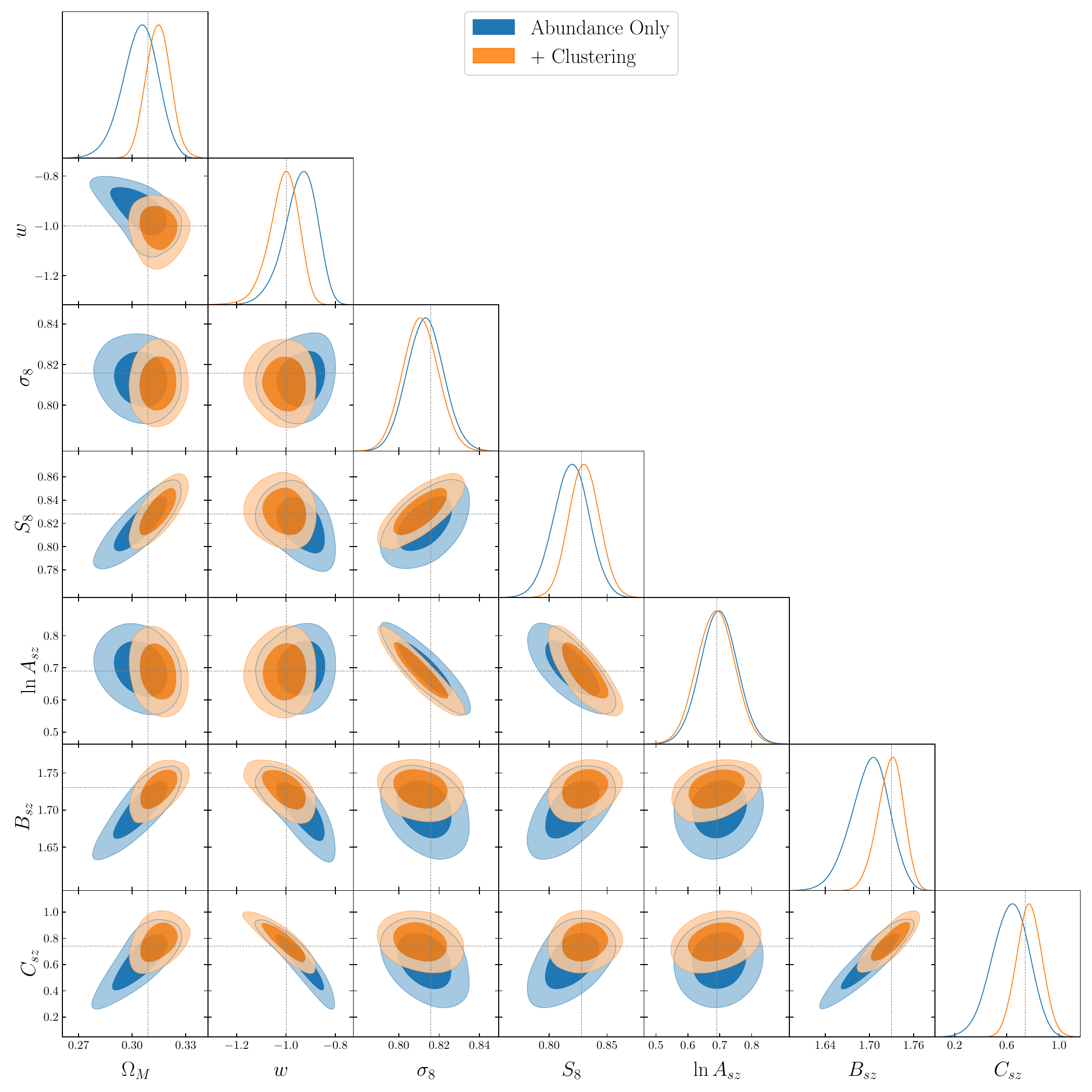}
\caption{1- and 2-$\sigma$  contours for a simulated 140k cluster sample with pessimistic scaling relation priors. 
The posteriors with from the cluster abundance measurement are shown in blue, while the posteriors with the measurement of the clustering power spectrum added are shown in orange. 
Clustering information allows a better determination of $B_{sz}$ and $C_{sz}$. 
This partially breaks the degeneracy between these parameters, $w$ and $\Omega_M$, thereby tightening the final measurements of $w$ and $\Omega_M$. 
\label{fig: degeneracies}}
\end{figure*}

To understand these changes to the 1d posteriors of $w$ and $\Omega_M$, we look at the parameter degeneracy structure, which is illustrated by the triangle plot shown in Figure~\ref{fig: degeneracies}.  
The figure shows the 1 and $2\,\sigma$ contours for the sampled cosmological parameters ($\Omega_M, w, S_8$), the derived cosmological parameter $\sigma_8$, and the sampled scaling relation parameters $\ln A_{sz}, B_{sz}, C_{sz}$. 
The results from cluster abundances alone are plotted in blue, while orange shows the posteriors for the combination of number counts and clustering. 
This figure is for the largest (140k) cluster sample with the pessimistic prior case to better illustrate the parameters degeneracy structure and the influence of clustering information.  

We first look at the ability of clustering information to inform the values of the three scaling relation parameters, $\ln A_{sz}, B_{sz}, C_{sz}$. 
From Figure~\ref{fig: degeneracies} and the final rows of Table \ref{tab:improvements}, we observe the inclusion of the clustering power spectrum measurement leads to sizeable improvements in constraints on the mass and redshift scaling parameters of the mass-observable relation, but has negligible impact on the overall mass normalization $\ln A_{sz}$. 
Under the pessimistic priors, clustering information significantly improves the determination of the mass slope $B_{sz}$ by a factor of $1.114\pm 0.015$ for the smallest, 33k cluster sample, which increases to $1.389\pm 0.041$ for the largest 140k cluster sample. 
Clustering information also provides comparable fractional improvements on the uncertainty of the redshift evolution parameter $C_{sz}$, by factors of $1.129\pm 0.021$ and $1.340\pm 0.039$ respectively for the smallest and largest samples. 
For the largest cluster sample, the observed reductions in uncertainties by factors of 1.34-1.39 suggest clustering information is providing nearly equal information as the pessimistic priors plus number counts on the mass slope $B_{sz}$ and redshift evolution $C_{sz}$ terms.

We next consider how a better determination of the mass-observable relation propagates to the cosmological constraints by examining the 2d contours between the scaling relation and cosmology parameters in   Figure~\ref{fig: degeneracies}. 
We start with the matter density $\Omega_M$. 
With number counts alone, we see a clear degeneracy between $\Omega_M$ and both the mass slope $B_{sz}$ and redshift evolution $C_{sz}$. 
Higher values of  $B_{sz}$ or $C_{sz}$ increase the matter density. 
There is little evidence for a degeneracy between the matter density and amplitude $\ln A_{sz}$. 
The $B_{sz}/C_{sz}$ degeneracies are largely broken with the better determination of these parameters enabled by the addition of clustering information, which significantly tightens the constraint on the matter density, as seen in Section~\ref{sec: Cosmological Parameter}. 

Turning to the dark energy equation of state parameter $w$, we find a very similar picture. 
The degeneracy between $w$ and the mass slope $B_{sz}$ and redshift evolution $C_{sz}$ has the opposite sign: higher values of  $B_{sz}$ or $C_{sz}$ decrease $w$. 
This degeneracy persists with the addition of clustering information, however the reduced range allowed for $B_{sz}$ and $C_{sz}$ still sharpens the determination of the dark energy equation of state.

In contrast to $w$ or $\Omega_M$, we see clustering information makes little change in the determination of $\sigma_8$.
For cluster abundance measurements, the main limit in determining $\sigma_8$ is the strong degeneracy between $\sigma_8$ and $\ln A_{sz}$. 
The additional clusters that are expected at higher $\sigma_8$ can be corrected for by lowering $\ln A_{sz}$ which effectively raises the mass threshold for detection.
Higher values of $\sigma_8$ also increase the predicted amplitude of the clustering power spectra. 
However the predicted power spectra only scales with $\sigma_8$ to a power of 2 to 5, while the predicted number counts scale with $\sigma_8$ to the power of 7 to 11. 
This leaves the clustering power spectra significantly weaker at constraining $\sigma_8$ than the abundance data with prior. 
The likelihood for the clustering data changes very little when moving $2\,\sigma$ along the major axis of the  $\sigma_8$-$\ln A_{sz}$ degeneracy ellipse.
For instance, shifting from the central value of $\ln A_{sz} = 0.6935$ and $\sigma_8 = 0.8151$ to $\ln A_{sz} = 0.8031$ and $\sigma_8 = 0.7996$ only results in a $\Delta \chi^2 = 0.3$ for the largest cluster sample. 
This drops to a negligible $\Delta \chi^2 = 0.05$ for the smallest cluster sample. 
Thus, clustering information has negligible impact on the 1D posterior for $\sigma_8$, although it slightly narrows the minor axes of the parameter ellipses involving $\sigma_8$.

 Stronger external priors fill a similar role as clustering information in breaking these degeneracies between the mass-observable parameters and cosmological parameters. 
 In the optimistic prior case, the higher-precision prior determines $B_{sz}$ and $C_{sz}$ better than clustering measurements, effectively breaking these parameter degeneracies without needing clustering data. 
 Consequently, we see that including clustering information does not significantly sharpen the posteriors on wCDM cosmological parameters when combined with the optimistic priors. 
 Still, even with good weak-lensing and other mass information, clustering information can still be used as an independent cross-check on the measured mass slope and redshift evolution.

\section{Conclusion}

We present forecasts for cosmological inferences within a wCDM cosmology for the combination of galaxy cluster abundance and clustering  measurements from upcoming tSZ cluster surveys. Clustering information is found to reduce the allowed parameter volume by tens of percent, with the largest improvements to $\Omega_M$ and $w$. 
We show that galaxy cluster clustering measurements provide an  independent cross-check on the redshift evolution and mass slope of the tSZ mass-observable scaling relation, without requiring extra observational power. 
The improvements in parameter inferences from the addition of clustering information increase for larger cluster samples and weaker external priors on the tSZ mass-observable scaling relation.

For the 140k cluster sample under pessimistic external mass priors, adding a measurement of the clustering of galaxy clusters to the number count data reduces the posterior width of the mass slope $B_{sz}$ and redshift evolution $C_{sz}$ of the tSZ mass-observable scaling relation by factors of $1.389 \pm 0.041$ and $1.340 \pm 0.039$ respectively. 
These significant reductions imply that galaxy cluster clustering measurements provide comparable information on the mass slope and redshift evolution parameters as the cluster number counts with pessimistic priors alone. 
These results highlight the potential value of self-calibration as an internal, independent cross-check on the mass slope and redshift evolution of the scaling relation. 

The improved determination of the tSZ mass-observable scaling relation enabled by the clustering measurement can lead to significant improvements in constraints on the matter density and dark energy equation of state, with the exact level dependent on the sample size and quality of the external mass information assumed. 
For the largest 140k galaxy cluster sample and baseline prior, the uncertainty on $w$ is reduced by a factor of $1.079\pm0.011$.  
The improvement on the matter density is larger still, with $\sigma(\Omega_M)$ reduced by a factor of $1.145\pm0.012$. 
The improvements fall substantially to $1.023\pm0.007$ and $1.068\pm0.007$ for $\sigma(w)$ and $\sigma(\Omega_M)$ respectively, when using the smallest sample (33k clusters) considered. 
However, even under tighter, optimistic mass priors, approximately 60\% of clustering information gains on $\sigma(w)$ remain.
By fully using all data from the tSZ galaxy cluster surveys, clustering measurements will help maximise the scientific return of upcoming tSZ cluster catalogues from experiments like SPT-3G and the Simons Observatory.

\section*{Acknowledgments}
This research is supported by The University of Melbourne's Research Computing Services and the Petascale Campus Initiative.

\bibliographystyle{apj}
\bibliography{biblio.bib}

\end{document}